\documentclass[preprint,showpacs,amsmath,amssymb,prb]{revtex4}
\usepackage{epsfig}
\begin{document}
\title{Density of states of a graphene in the presence of strong point defects}

\author{Bor-Luen Huang$^{1}$, Ming-Che Chang$^{1}$, and Chung-Yu Mou$^{2,3,4}$}
\affiliation{
1. Department of Physics, National Taiwan Normal University, Taipei, Taiwan\\
2. Department of Physics, National Tsing Hua University, Hsinchu, Taiwan\\
3. Institute of Physics, Academia Sinica, Nankang, Taiwan\\ 4.
Physics Division, National Center for Theoretical Sciences, P.O.Box
2-131, Hsinchu, Taiwan}

\date{\today}

\begin{abstract}
The density of states near zero energy in a graphene due to strong
point defects with random positions are computed. Instead of
focusing on density of states directly, we analyze eigenfunctions of
inverse T-matrix in the unitary limit. Based on numerical
simulations, we find that the squared magnitudes of eigenfunctions
for the inverse T-matrix show random-walk behavior on defect
positions. As a result, squared magnitudes of eigenfunctions have
equal {\it a priori} probabilities, which further implies that the
density of states is characterized by the well-known Thomas-Porter
type distribution. The numerical findings of Thomas-Porter type
distribution is further derived in the saddle-point limit of the
corresponding replica field theory of inverse T-matrix. Furthermore,
the influences of the Thomas-Porter distribution on magnetic and
transport properties of a graphene, due to its divergence near zero
energy, are also examined.
\end{abstract}


\maketitle
\section{Introduction}
Recently, the isolation of single-layer graphene\cite{Geim07} has
revived much interest in studying two-dimensional (2D) Dirac
fermions. One of the peculiar properties associated with 2D Dirac
fermions is the unusual electronic properties in the presence of
defects and disorders. In the context of cuprate superconductors,
where quasi-particles are also 2D Dirac fermions, disorders have
masked the d-wave nature and hindered its discovery. It was later
realized that point defects may change the density of states (DOS)
near the Dirac point and strong point defects may even induce
quasi-localized states or magnetic moments near zero energy in
d-wave superconductors\cite{Balatsky}. In the case of graphene, it
is found that there is finite density of states due to weak
disorders\cite{Tan}. For strong disorders, it was observed that
ferromagnetic state can be induced by bombarding a graphite with
protons\cite{Esquinazi}. The induced magnetism is further confirmed
to be resulted from $\pi$-electrons\cite{Ohldag}. This fact,
together with recent observation of ferromagnetism in disordered
graphene\cite{Wang,Ramakrishna}, shows that graphene with defects
could become ferromagnetic. In addition to magnetism, graphene also
reveals anomalous transport properties in the presence of strong
disorders, where in the presence of vacancies, instead of
decreasing, the conductivity is found to increase\cite{Coleman}.
These observations clearly indicates that in the presence of strong
disorders, 2D Dirac fermions may behave very differently from what
is expected for clean or weak disordered graphene.

Experimentally, there are many possible forms of disorders in
graphene\cite{Hashimoto}. For large defects such as cracks, they
tend to contain the so-called zig-zag edges, where localized
states would appear near the edge\cite{Mou} and induce magnetic
behavior\cite{Hikihara}. In this case, magnetic moments arise from
localized states and interact via RKKY
(Ruderman-Kittel-Kasuya-Yosida) interaction, which tends to make
graphene antiferromagnetic\cite{Brey}. Hence the most possible
candidates for the observed ferromagnetism in graphene are defects
of small sizes or simply point defects. Here the simplest point
defects are single-atom vacancies or hydrogen chemisorption
defects. These kinds of defects generally create complicated
disturbances in graphene and may even form ordered
structures\cite{McCann}. However, for low density of quenched
defects, they can be simulated by a large potential $u$ on a
lattice point without distortion of nearby lattice
points\cite{yaz}.

Theoretically, extensive studies on a single defect have been
performed on d-wave superconductors\cite{Balatsky}. It is known that
a zero-energy electronic state would arise near a point defect with
$u \rightarrow \infty$ or a circular disk in a 2D Dirac
Hamiltonian\cite{Dong}. Furthermore, the electronic wavefunction is
semi-localized with amplitude decaying as $1/r$ at large distance
$r$\cite{Balatsky, Pereira,Dong}. The semi-localized behavior is
clearly revealed in the observed STM images of long-range $(\sqrt{3}
\times \sqrt{3}) R 30^{\circ}$ superstructure in
graphene\cite{STMAFM,Ugeda}. For finite density of defects, one
expects that semi-localized electrons interact strongly and may form
an impurity band\cite{Balatsky,mou09}. Nonetheless, conflicting
results based on either perturbative or non-perturbative approaches
are reported\cite{Balatsky}. The residual DOS near zero energy is
predicted to be either finite\cite{Ziegler},
infinite\cite{Lee,Ting}, or vanishing with different power laws in
energy\cite{Balatsky}. This issue remains unsolved.

While quasi-particles in both cuprate superconductors and graphenes
are 2D Dirac fermions, the situation is quite different for
graphene. For neutral graphene, even though excitonic effects are
expected to be large\cite{Louie}, for low energies and large
distances, the screened Coulomb interaction is shown be
long-ranged\cite{Sheehy} with renormalized dielectric constant.
Furthermore, the electron itself is the quasi-particle and carries a
definite charge. These differences make graphene behave totally
different from that of cuprate superconductors in the strong
disorders. In particular, without being masked by superconductivity,
direct manifestation of the impurity band is possible in graphene.
Therefore, investigation on graphene with strong defects would
provide an unique opportunity to clarify the issue of DOS near zero
energy for 2D Dirac fermions with strong disorders. This is recently
pointed out in Ref.[\onlinecite{mou09}].  In that paper, the
wavefunction for finite density of defects is constructed. By using
the wavefunction for two defects, it is shown that ferromagnetic
state is favored for large distances between two defects. However,
for finite density of defects, the problem of finding DOS is mapped
to an equivalent problem of finding the DOS of a random matrix. One
has to assume that the matrix elements are independent random
numbers to demonstrate the induced ferromagnetism\cite{mou09}. While
the predicted DOS (Wigner semi-circle law) appears to be consistent
with results obtained by self-consistent Born
approximation\cite{Peres}, to confirm that the observed
ferromagnetism and anomalous transport properties of graphene are
consequences of the impurity band, one needs to go beyond
self-consistent Born approximation and to resolve the issue of how
the DOS of 2D Dirac fermions changes in strong disorder limit.

In this paper, we re-examine the density of states of a graphene due
to strong point defects. In particular, we show that the inverse
T-matrix for $N_I$ point defects can be exactly mapped to a $N_I
\times N_I$ symmetric Euclidean Random Matrix in which one cannot
treat the matrix elements as independent random numbers. Instead of
focusing on the DOS directly, we analyze magnitudes distribution for
eigenfunctions for the derived Random Matrix. Remarkably, we find
that squared magnitudes of eigenfunctions show random-walk behaviors
on defect positions. As a result, the distribution of squared
magnitudes of eigenfunctions for the Euclidean Random Matrix follows
the Porter-Thomas distribution. Further analysis shows that
eigenvalues ($\lambda$) of the corresponding Euclidean Random Matrix
also follow the Thomas-Porter distribution\cite{ThomasPorter} and
the DOS near zero energy for infinite $u$ is
\begin{equation}
D(E) =  n_I  \sqrt{\frac{1}{8 \pi \langle |\lambda| \rangle |\lambda
(E)| }}e^{- \frac{|\lambda (E) |}{2 \langle |\lambda| \rangle} }
\left| \frac{ \lambda(E) }{dE} \right|. \label{DOS0}
\end{equation}
Here $n_I$ is the density of defects and $\langle |\lambda| \rangle
$ is the average of $|\lambda|$ over defect configurations. $\lambda
(E)$ is given by $\lambda (E) = - \frac{\sqrt{3}}{2 \pi D^2}E \ln
|E/D|$ with $D=3t/2$ and $t$ being the hopping amplitude of the
electron. This form of the density of states is valid when $|E| \ll
t$ and we found that $\langle |\lambda| \rangle  \sim \sqrt{n_I}$
shows random-walk behavior. The resulting density of states has
strong effects on magnetic and transport properties of graphene. We
re-examine the effect of the long-range Coulomb interaction with
renormalized dielectric constant and show that the resulted DOS
supports ferromagnetism for any finite density of defects. At finite
temperature, the linear extrapolation of magnetization curve
indicates that $T_c \sim 600-700 K$, in agreement with experimental
observations.

This paper is organized as follows. In Sec. II, we lay down the
theoretical formulation and show that the inverse T-matrix for $N_I$
point defects can be exactly mapped to a $N_I \times N_I$ Euclidean
Random Matrix. In Sec. III, we use both analytic arguments and
numerical simulations to derive the density of resonant states. In
Sec. IV, we reexamine effects of the screened long-range Coulomb
interaction. We show that the competition between the exchange
energy and kinetic resonant energy leads to ferromagnetism for
infinite on-site potentials. The magnetizations both at zero and
finite temperatures are also calculated. In Sec. V, we conclude and
discuss possible effects for weak impurities. Appendix A is devoted
to more rigorous derivation of the Porter-Thomas distribution in the
saddle-point limit.

\section{Theoretical Formulation } \label{formulation}
\begin{figure}[h]
\includegraphics[width=4.5cm]{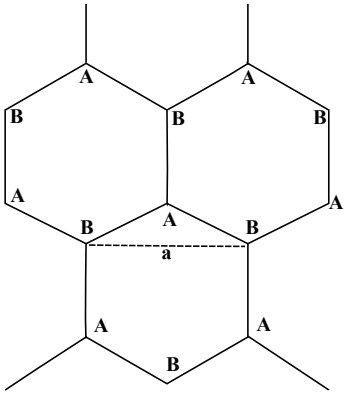} \caption{Honeycomb
lattice of graphene. The lattice is bi-partite, labeled by A and B ,
with hopping amplitude between nearest A and B being $t \sim$2.7eV.
The lattice constant $a=2.46 \AA$ is the distance between two
nearest B points.} \label{graphene}
\end{figure}

We start by setting up the framework for investigating the effects
of defect. It is known that electrons in the $\pi$ band of an
infinite graphene can be well described by a tight-binding
Hamiltonian $H_0$\cite{Geim07}. As shown in Fig. \ref{graphene},
the lattice of graphene is bi-partite. If we label the bi-partite
lattice points by A and B, $H_0$ consists of hopping only for
nearest A and B with a hopping amplitude $t$. Hence if defects are
located at $\vec{r}_i$ with $i=1,2,3, \cdots, N_I$, the
wavefunction $\psi $ for an electron then satisfies
\begin{equation}
\left( H_{0} + u \sum^{N_I}_{i=1} \delta_{\vec{r},\vec{r}_i} \right)
\psi_E ({\vec{r}}) = E \psi_E ({\vec{r}}). \label{wavefunction}
\end{equation}
Here and in the following, both $\vec{r}$ and $\vec{r}_i$ are
restricted to points on the honeycomb lattice shown in Fig. 1.
To find the effects of defects on the electronic state, it is
sufficient to calculate the Green's function $G
({\vec{r},\vec{r}', E}) $, which describes the amplitude for the
electron to propagate from $\vec{r}'$ to $\vec{r}$ and satisfies
\begin{equation}
\left(E- H \right)  G ({\vec{r},\vec{r}', E}) =
\delta_{\vec{r},\vec{r}'},  \label{Geq}
\end{equation}
where $H=H_0 +  u \sum^{N_I}_{i=1} \delta_{\vec{r},\vec{r}_i}$.
For clean graphene, the Green's function will be denoted by $G^0
({\vec{r},\vec{r}', E}) $. In the Fourier $k$ space, it is
convenient to reorganize the wavefunction into $\psi_A$ and
$\psi_B$ for A and B sublattices. Then $G^0(k)$ is the inverse of
the $2 \times 2$ matrix, $E+ i 0^+ - H_0(k)$, with $H_0(k)$ being
given by
\begin{equation}
H_0(k) = \left(
\begin{array}{cc}
0 & \Delta (k) \\
\Delta^* (k) & 0
\end{array}
\right),
\end{equation}
where $\Delta (k) = - t[2 e^{ik_y a2\sqrt{3}} \cos
(k_xa/2)+e^{-ik_ya/\sqrt{3}} ]$. More explicitly, one finds
\begin{eqnarray}
&& G^0_{AA}=G^0_{BB} \nonumber \\
&& = \frac{1}{2} \left[ \frac{1}{E+i0^+ -|\Delta(k)|} +
\frac{1}{E+i0^+ +|\Delta(k)|} \right] \label{GAA} \\
&& G^0_{AB}  \nonumber \\
&&= \frac{\Delta (k) }{2 |\Delta (k)| } \left[ \frac{1}{E+i0^+
-|\Delta(k)|} - \frac{1}{E+i0^+ +|\Delta(k)|} \right] \nonumber \\
&& G^0_{BA} \nonumber \\
&& = \frac{\Delta^* (k) }{2 |\Delta (k)| } \left[ \frac{1}{E+i0^+
-|\Delta(k)|} - \frac{1}{E+i0^+ +|\Delta(k)|} \right] \nonumber
\\ \label{GBA}
\end{eqnarray}
In real space, it is more convenient to use lattice vectors
$\vec{r}$ to carry indices for A and B sublattice. Therefore, $G$ is
no longer a $2 \times 2$ matrix and can be expressed in terms of
$G^0$ as
\begin{equation}
 G({\vec{r},\vec{r}', E})= G^0({\vec{r},\vec{r}', E}) + u
 \sum^{N_I}_{i=1} G^0({\vec{r},\vec{r}_i, E}) G({\vec{r}_i,\vec{r}',
 E}). \label{sumperturb}
\end{equation}
Clearly, to find $G ({\vec{r},\vec{r}', E}) $, one needs to find
$G({\vec{r}_i,\vec{r}', E})$ in Eq. (\ref{sumperturb}). For this
purpose, one sets $\vec{r}$ to $\vec{r}_i$ with $i=1,2,3, \cdots,
N_I$ in Eq. (\ref{sumperturb}) and solves $G({\vec{r}_i,\vec{r}',
E})$ in terms of $G^0({\vec{r}_i,\vec{r}'})$. If we replace the
notation $G({\vec{r}_i,\vec{r}', E})$ by $G_{{\vec{r}_i,\vec{r}'}}$
with $E$ being suppressed,  we obtain
\begin{widetext}
\begin{eqnarray}
\left(
\begin{array}{c}
G_{{\vec{r}_1,\vec{r}'}} \\
G_{{\vec{r}_2,\vec{r}'}} \\
G_{{\vec{r}_3,\vec{r}'}}  \\
\cdot    \\
\cdot    \\
G_{{\vec{r}_{N_I},\vec{r}'}}
\end{array}
\right) =  1/u \left(
\begin{array}{cccccc}
1/u-G^0_{{\vec{r}_1,\vec{r}_1}} & -G^0_{{\vec{r}_1,\vec{r}_2}} & -G^0_{{\vec{r}_1,\vec{r}_3}} & \cdot  & \cdot  & -G^0_{{\vec{r}_1,\vec{r}_{N_I}}} \\
-G^0_{{\vec{r}_2,\vec{r}_1}} & 1/u-G^0_{{\vec{r}_2,\vec{r}_2}} & -G^0_{{\vec{r}_2,\vec{r}_3}} & \cdot  & \cdot  & -G^0_{{\vec{r}_2,\vec{r}_{N_I}}} \\
-G^0_{{\vec{r}_3,\vec{r}_1}} & -G^0_{{\vec{r}_3,\vec{r}_2}} & 1/u-G^0_{{\vec{r}_3,\vec{r}_3}} & \cdot  & \cdot  & -G^0_{{\vec{r}_3,\vec{r}_{N_I}}} \\
\cdot  & \cdot  & \cdot  & \cdot  & \cdot  \\
\cdot  & \cdot  & \cdot  & \cdot  & \cdot  \\
-G^0_{{\vec{r}_{N_I},\vec{r}_1}} & -G^0_{{\vec{r}_{N_I},\vec{r}_2}}
& -G^0_{{\vec{r}_{N_I},\vec{r}_3}} & \cdot & \cdot  &
1/u-G^0_{{\vec{r}_{N_I},\vec{r}_{N_I}}}
\end{array}
\right)^{-1}
 \left(
\begin{array}{c}
G^0_{{\vec{r}_1,\vec{r}'}} \\
G^0_{{\vec{r}_2,\vec{r}'}} \\
G^0_{{\vec{r}_3,\vec{r}'}}  \\
\cdot    \\
\cdot    \\
G^0_{{\vec{r}_{N_I},\vec{r}'}}
\end{array}
\right) . \label{T} \end{eqnarray} \end{widetext} Here the matrix on
the right hand side is the T-matrix whose inverse determines
resonant energies and can be separated into real and imaginary parts
\begin{widetext}
\begin{eqnarray}
T^{-1}= \left(
\begin{array}{ccccc}
1/u-\cal{G}_{\it{11}} & -\cal{G} _{\it{12}} &  \cdot  & \cdot  & -\cal{G}_{\it{1N_I}} \\
-\cal{G}_{\it{21}} & 1/u-\cal{G}_{\it{22}} &  \cdot  & \cdot  & -\cal{G}_{\it{2N_I}} \\
-\cal{G}_{\it{31}} & -\cal{G}_{\it{32}} & \cdot  & \cdot  & -\cal{G}_{\it{3N_I}} \\
\cdot  & \cdot  & \cdot  & \cdot  & \cdot  \\
\cdot  & \cdot  & \cdot  & \cdot  & \cdot  \\
-\cal{G}_{\it{N_I1}} & -\cal{G}_{\it{N_I2}} & \cdot  & \cdot  &
1/u-\cal{G}_{\it{N_I N_I}}
\end{array}
\right) - i \left(
\begin{array}{ccccc}
I_{11} & I_{12} &  \cdot  & \cdot  & I_{1N_I} \\
I_{21} & I_{22} &  \cdot  & \cdot  & I_{2N_I} \\
I_{31} & I_{32} & \cdot  & \cdot  & I_{3N_I} \\
\cdot  & \cdot  & \cdot  & \cdot  & \cdot  \\
\cdot  & \cdot  & \cdot  & \cdot  & \cdot  \\
I_{N_I1} & I_{N_I2} & \cdot  & \cdot  & I_{N_I N_I}
\end{array}
\right)
 , \label{eigen} \end{eqnarray}\end{widetext}
where $\cal{G}_{\it{ij}}$ and $I_{ij}$ are the real and imaginary
parts of $G^0_{ij}$. Note that due to the Kramers-Kronig relation,
$\cal{G}_{\it{ij}}$ and $I_{ij}$ are related by
\begin{equation}
I_{ij} (E)  = \cal{P} \it{ \int \frac{\cal{G}_{\it{ij}}(
\it{E'})}{E-E'} dE'}. \label{Kramers}
\end{equation}
Therefore, real ($T^{-1}_R$) and imaginary parts ($T^{-1}_I$) of
$T^{-1}$ can be diagonalized simultaneously. In particular, their
eigenvalues are also related by the Kramers-Kronig relation
\begin{equation}
\lambda_I (E)  = \cal{P} \it{ \int \frac{\lambda_G(E')}{E-E'} dE'}.
\label{Kramers_lamda}
\end{equation}
It is thus clear that the resonant energies of the Green's function
$G$ are determined by zeros of eigenvalues of $T_R$. Therefore,
resonant energies due to defects are determined by
\begin{equation}
\left|
\begin{array}{ccccc}
1/u-\cal{G}_{\it{11}} & -\cal{G} _{\it{12}} &  \cdot  & \cdot  & -\cal{G}_{\it{1N_I}} \\
-\cal{G}_{\it{21}} & 1/u-\cal{G}_{\it{22}} &  \cdot  & \cdot  & -\cal{G}_{\it{2N_I}} \\
-\cal{G}_{\it{31}} & -\cal{G}_{\it{32}} & \cdot  & \cdot  & -\cal{G}_{\it{3N_I}} \\
\cdot  & \cdot  & \cdot  & \cdot  & \cdot  \\
\cdot  & \cdot  & \cdot  & \cdot  & \cdot  \\
-\cal{G}_{\it{N_I1}} & -\cal{G}_{\it{N_I2}} & \cdot  & \cdot  &
1/u-\cal{G}_{\it{N_I N_I}}
\end{array}
\right| =0. \label{eigen} \end{equation} Note that the above
condition is exactly the same as the one obtained via the
constructed wavefunction for defects\cite{mou09} and should be
compared to the similar equation obtained in the context of d-wave
superconductors\cite{Lee}. Since a graphene without defect is
translationally invariant, one has $G^0_{ij}=G^0(\vec{r}_i
-\vec{r}_j)$. Therefore, diagonal terms in Eq.(\ref{eigen}) are
identical and are equal to $\lambda(E) \equiv 1/u - \rm{Re} \it{
G^0} (\rm{0},\it{E})$. Hence if the positions of defects are
random, solving Eq.(\ref{eigen}) is equivalent to finding
eigenvalues of the random matrix
\begin{equation}
H_I = \it{ \left(
\begin{array}{ccccc}
0 & \cal{G} _{\it{12}} &  \cdot  & \cdot  & \cal{G}_{\it{1N_I}} \\
\cal{G}_{\it{21}} & 0 &  \cdot  & \cdot  & \cal{G}_{\it{2N_I}} \\
\cal{G}_{\it{31}} & \cal{G}_{\it{32}} & \cdot  & \cdot  & \cal{G}_{\it{3N_I}} \\
\cdot  & \cdot  & \cdot  & \cdot  & \cdot  \\
\cdot  & \cdot  & \cdot  & \cdot  & \cdot  \\
\cal{G}_{\it{N_I1}} & \cal{G}_{\it{N_I2}} & \cdot  & \cdot  & 0
\end{array}
\right) }.  \label{random} \end{equation} Furthermore, if one
defines the density of eigenvalues for $H_I$ by
\begin{equation}
{\cal D}(\lambda) = \frac{1}{M} \sum_n \delta (\lambda - \lambda_n),
\end{equation}
with $M$ being the total number of lattice points and $\lambda_n$
being the $n$-th eigenvalue, the density of resonant states is
given by
\begin{equation}
D(E) =  \cal{D}(\it{\lambda (E) }) \left| \frac{ d \lambda(E) }{dE}
\right|.
\end{equation}
Therefore, it is sufficient to find the distribution of eigenvalues
for $H_I$. We note in passing that if values of $\lambda_G$ form a
band after averaging over defect configurations, it implies that the
averaged $\langle \lambda_G \rangle$ is independent of $E$.
Eq.(\ref{Kramers_lamda}) then implies that except for contributions
from diagonal terms $I_{nn}$, off-diagonal terms do not contribute
to the imaginary part of eigenvalues. Hence if values of $\lambda$
form a band, one has $T^{-1}_I= - \rm{Im} \it{ G^0} (\rm{0},\it{E})
\bf{I} $. Since $\rm{Im} \it{ G^0} (\rm{0},\it{E}) \propto E$, this
result implies that the inverse of lifetime for resonant states is
proportional to $E$, consistent with experimental
observation\cite{lifetime}.

\section{Density of resonant states} \label{dos}
In the last section, it is shown that the density of resonant
states is determined by the spectrum of $H_I$. Since each element,
$\cal{G}_{\it{ij}}$, depends on positions of defects, they
fluctuate randomly. In the simplest approximation, one treats each
element as an independent random number. The density of states is
characterized by the Wigner semi-circle law\cite{mou09}. As
indicated earlier, this approximation appears to be equivalent to
the self-consistent Born approximation \cite{Peres}. A closer
examination of $H_I$ shows that the dependence of each matrix
element on the position $\vec{r}_i$ makes them correlated. Hence
one cannot treat each element as an independent random number.
Indeed, it was realized in different context by Mezard et
al.\cite{Zee} that such random matrices form distinct classes
known as Euclidean Random Matrices, whose spectrum depends on the
functional form of the matrix element on $\vec{r}_i$ .

It is generally difficult to find the exact spectrum for any given
Euclidean Random Matrices. For defects on graphene, however, it
turns out that the spectrum of $H_I$ follows a simple form known
as the Porter-Thomas distribution\cite{ThomasPorter}. In this
section, we shall focus on the study of the spectrum by numerical
simulation. An analytical derivation based on saddle-point
approximation will be relayed to the Appendix.

We start by noting that since one expects that the energies of
resonant states are close to zero, as a first step, we can
approximate each matrix element by $\cal{G}_{\it{ij}} \it(E=0)$.
We shall see later that the error due to this approximation is
small for $E \sim 0$. In this approximation, by using
Eqs.(\ref{GAA}) and (\ref{GBA}), one finds ${\cal
G}_{AA}(r,E=0)={\cal G}_{BB} (r,E=0)=0$ and $G_{BA}(j,i) =
G^*_{AB} (i,j)$. Hence $H_I$ is a symmetric matrix. Furthermore,
since in the second quantization form, $H_I = \sum_{ij} {\cal
G}_{ij} {c^{\dagger}_A}_i {c_B}_j + h.c.$, we find that $H_I$ goes
to $-H_I$ under the particle-hole transformation:
${c^{\dagger}_A}_i \rightarrow - {c_A}_i$ and ${c_B}_j \rightarrow
{c^{\dagger}_B}_j$. Therefore, the spectrum is particle-hole
symmetric, i.e., ${\cal D} (-\lambda) = {\cal D} (\lambda)$. In
addition of being particle-hole symmetric, $H_I$ itself also
supports energy states exactly at zero energy due to the unbalance
in the number of lattice points in A and B\cite{Brouwer}. Since
the number of zero energy states is equal to $|N_A -N_B|$, if
lattice points are randomly assigned to A or B, one finds $|N_A
-N_B| \sim \sqrt{N_I}$ and hence their contribution is negligible
in the limit of $M \rightarrow \infty$ with $N_I/M$ being fixed at
the defect density $n_I$. Therefore, in the following, we shall
focus on density of resonant states for the case with $N_A = N_B$
to avoid complications due to extra zero energy states.

For high density of defects, because the positions of defects
sample sufficient lattice points, $H_I$ can be diagonalized by
Fourier transformation. Hence eigenvalues of $H_I$ are
proportional to the Fourier transformation of $\cal{G}_{\it{ij}}$.
We find that
\begin{eqnarray}
{\cal D} (\lambda) \propto \int \int \frac{ d^2 {\bf q}}{(2 \pi)^2}
\left[ \delta \left( \lambda - \frac{1}{|\Delta({\bf q})|} \right) +
\delta \left( \lambda + \frac{1}{|\Delta({\bf q})|} \right)\right].
\nonumber \\
\label{highdensity}
\end{eqnarray}
In this case, because $0 \leq |\Delta(q)| \leq 3t $, we obtain
$\lambda \geq 1/3t $. Therefore, there is no resonant defect state
near zero energy for sufficient high density of defects.

\begin{figure}[htbp]
\hspace{-0.2in} \includegraphics[width=7.5cm]{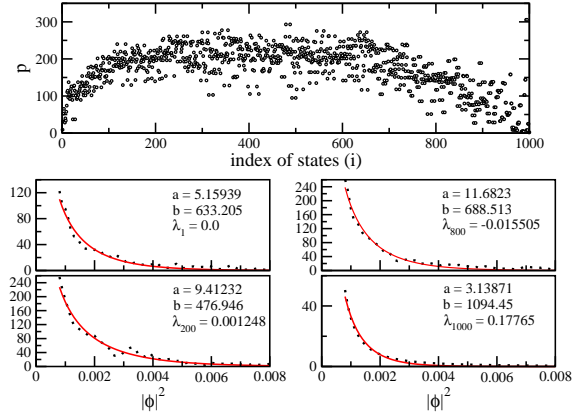}
\caption{(Color on-line) Participation number ($p = 1/ \sum_i
|\phi(\vec{r}_i)|^4$) and histogram of $|\phi_{\lambda}|^2$ of
different $\lambda$'s and $p$'s (indicated by sub-indices of
$\lambda$) for a typical defect configuration simulated with
$N_I=1000$ and $M=1000 \times 1000$, i.e., $n_I = 0.001$. Here red
solid lines are the fitted Boltzmann distributions.} \label{TP1}
\end{figure}
For low density of defects, the separation between any two defects
is large. In this case, by using Eqs.(\ref{GAA}) and (\ref{GBA}), we
find that for $0<|E| r/v \ll 1$\cite{Lee},
\begin{eqnarray}
&& {\cal G}_{AA}(r,E)= {\cal G}_{BB}(r,E) = \frac{\sqrt{3} a^2}{2
\pi v^2} \cos \left(
\frac{4\pi x}{3a}  \right) E \ln \frac{r|E|}{v}, \nonumber \\
&& \label{GAAr} \\
&& {\cal G}_{AB}(r,E) = {\cal G}_{BA}(r,E) = \frac{\sqrt{3} a}{2 \pi
v} \frac{1}{r/a} \sin \left( \frac{4\pi x}{3a}  \right).
\label{GABr}
\end{eqnarray}
Here $v = 3ta/2$. While for $E=0$, as we indicated earlier, ${\cal
G}_{AA}(r,E=0)={\cal G}_{BB} (r,E=0)=0$ but ${\cal G}_{AB}(r,E=0)$
is given by Eq.(\ref{GABr}). For $r=0$, we obtain
\begin{equation}
{\cal G}_{AA}(0,E)= {\cal G}_{BB}(0,E) = \frac{\sqrt{3} a^2 }{2 \pi
v^2} E \ln \frac{a|E|}{v}. \label{AA0}
\end{equation}
To motivate it, instead of focusing on DOS directly as done in the
d-wave superconductors\cite{Lee,Balatsky}, we analyze the
distribution of the eigenfunction amplitudes $\phi_{\lambda}
(\vec{r}_i) $ of $H_I$ at a fixed eigenvalue $\lambda$
\begin{equation}
P(|\phi|^2) = \frac{1}{M} \sum_i \delta (|\phi|^2 - |\phi_{\lambda}
(\vec{r}_i)|^2).
\end{equation}
Here $\phi_{\lambda} (\vec{r}_i) $ is normalized so that
\begin{equation}
\sum_i |\phi_{\lambda} (\vec{r}_i)|^2 =1. \label{norm}
\end{equation} Hence if
there is no bias on partitioning $|\phi|^2$, one expects $|\phi|^2$
follows the Boltzmann type distribution, $P(|\phi|^2) \propto
e^{-\alpha |\phi|^2 }$. Indeed, in the limit $M \rightarrow \infty$,
Porter and Thomas\cite{ThomasPorter} derived the following
distribution
\begin{equation}
P(t) = \frac{1}{2 \pi s \langle s \rangle }  e^{- \frac{s}{2 \langle
s \rangle } }, \label{PT1}
\end{equation}
where $s = |\phi|^2$ and $\langle s \rangle$ is the average of
$|\phi|^2$. The same distribution can also be derived in the
non-linear sigma model\cite{Efetov1}. The Porter-Thomas
distribution, however, is not universal and is valid only when the
system is sufficiently chaotic\cite{Muller}. Since the matrix
element ${\cal G}_{ij}$ decays slowly ($1/r$), $\phi_{\lambda}$ at
each point $\vec{r}_i$ is determined by all other defects with
random positions. In other words, $H_I$ is a random hopping model
in which $\phi_i$ characterizes density of random walkers on
defect position $\vec{r}_i$. Since the probability for finding a
random walker at the traveling distance $r$ is proportional to
$e^{-r^2/2\langle r^2 \rangle}$, by comparison with
Eq.(\ref{PT1}), one expects that the Porter-Thomas distribution
works for $H_I$ with $\phi$ playing the role of distance. More
explicitly, for a random walker described by $\vec{r} (t)$, one
finds $ \langle r^2 \rangle \propto t$ at time $t$. Here $t$
characterizes the number of attempts in a random walk. By analogy,
$N_I$ would be the number of attempts. Therefore, we expect
\begin{equation}
\langle |\phi|^2 \rangle \propto \sqrt{n_I}. \label{randomwalk}
\end{equation}
Based on Eq.(\ref{GABr}), we perform extensive numerical analysis
on the statistics of eigenstates of $H_I$. To see if there is
correlation between distribution and localization of
$\phi_{\lambda}$, we also analyze the participation number $p = 1/
\sum_i |\phi(\vec{r}_i)|^4$ and find the distribution for
different participation numbers. Fig. \ref{TP1} shows the
statistics of wavefunction amplitudes for a typical defect
configuration. It is clear that regardless of whether the
eigenfunction is localized or not, distribution of amplitudes
follow the Porter-Thomas distribution for all participation
numbers.

For different $\lambda$, in addition to Eq.(\ref{norm}), partition
of eigenfunction amplitudes $\phi$ has an addition constraint
\begin{equation}
\sum_{ij} \phi_i (H_I)_{ij} \phi_j = \lambda, \label{lambda}
\end{equation}
where $(H_I)_{ij} = {\cal G}_{ij}$ for $i \neq j$ and $(H_I)_{ii} =
0$. It is clear that for different $\lambda$, $\phi \propto
\sqrt{\lambda}$. Hence by replacing $\phi$ by $\sqrt{|\lambda|}$ in
Eq. (\ref{PT1}) with appropriate normalization, we expect that the
distribution for $\lambda$ also follows the Porter-Thomas
distribution
\begin{equation}
{\cal D} (\lambda ) = n_I  \sqrt{\frac{1}{8 \pi \langle |\lambda|
\rangle |\lambda| }}e^{- \frac{|\lambda|}{2 \langle
|\lambda|\rangle} }. \label{Dlambda}
\end{equation}
Here according to Eq.(\ref{randomwalk}), we expect $\langle
|\lambda|\rangle \propto \sqrt{n_I}$. The proportional constant will
be determined numerically. The normalization ${\cal D}$ in
Eq.(\ref{Dlambda}) is chosen by requiring $\int^{\infty}_{-\infty} d
\lambda {\cal D} (\lambda) = n_I$. Note that for later use in the
calculation of magnetization, the normalization of ${\cal D}$ has to
be done by taking into account the presence of Dirac band.

\begin{figure}[htbp]
\hspace{-0.2in}
\includegraphics[width=7.5cm]{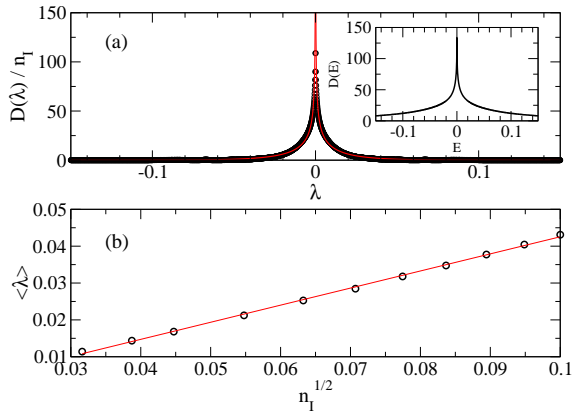}
\caption{(Color on-line) (a) Averaged spectrum of $H_I$ (with
${\cal G}_{ij} (E=0)$ as the matrix element) over 1000 defect
configurations. Here $n_I=1000$, $M=1000 \times 1000$ and we have
set $t=1$. Black circles are numerical results while the red line
is the fitted Porter-Thomas distribution with ${\cal D} (\lambda )
=2.17 e^{-45.438 |\lambda|} / \sqrt{|\lambda|}$. Inset: The
corresponding density of electronic states for $u = \infty$. (b)
Random-walk behavior of $\phi$: The dependence of $\langle \lambda
\rangle$ on the defect density $\sqrt{n_I}$ shows linear behavior
with a slope $0.464$. Here open circles are numerical results
obtained by the fitted Porter-Thomas distribution while the red
line is the linear curve of slope $0.464$. There is a small error
offset by $-0.0038$.} \label{TP2}
\end{figure}

Fig.\ref{TP2}(a) shows a typical spectrum of our numerical
simulations of the spectrum averaged over 1000 defect
configurations. It shows that the spectrum can be well described by
the Porter-Thomas distribution. Fig.\ref{TP2}(b) shows the fitted
parameter $\langle \lambda \rangle$ versus density of defects. It
indicates that $\langle \lambda \rangle$ follows a simple form of
$n_I$ by $\langle \lambda \rangle \approx \sqrt{n_I}$. Once one
knows the spectrum of $H_I$, by using Eq.(\ref{AA0}), the density
of resonant energies can be found by setting $\lambda = 1/u -
{\cal G}_{AA/BB} (0,E)$. This results in Eq.(\ref{DOS0}). In the
inset of Fig.\ref{TP2}(a), we show the corresponding electronic
DOS for $u=\infty$. It is clear that the DOS diverges at $E=0$.

\begin{figure}[htbp]
\hspace{-0.2in}
\includegraphics[width=7.0cm]{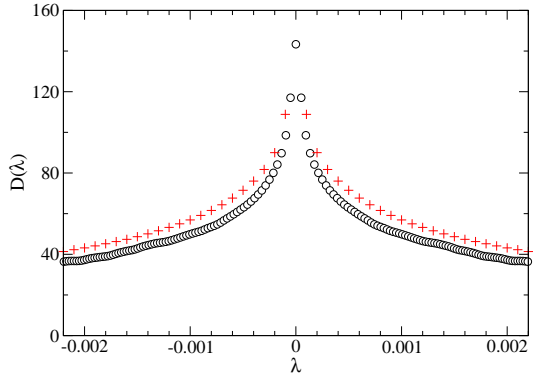}
\caption{(Color on-line) Comparison of the spectrum of $H_I$
determined by using ${\cal G}_{ij} (E)$ (open circles) and ${\cal
G}_{ij} (E=0)$ (red crosses) as matrix elements. Here $n_I=1000$,
$M=1000 \times 1000$ and we have set $t=1$. Both spectra can be
fitted with the Porter-Thomas distributions with slightly
different $\langle |\lambda| \rangle$.} \label{TP3}
\end{figure}

We close this section by checking the validity of setting $E=0$ in
${\cal G}_{ij} (E)$. For a given finite $E$, because $\lambda =
1/u - {\cal G}(0,E)$, there is only one value of $\lambda$
corresponding to the given $E$. Hence for a given $E$, only the
spectrum at $\lambda = 1/u - {\cal G}(0,E)$ is correct. To get the
whole spectrum, it is necessary to vary $E$ and obtain the
spectrum at each $\lambda (E)$ one by one. Note that by using
Eqs.(\ref{GAA}) and (\ref{GBA}), one finds that ${\cal G}_{ij}
(-E) = -{\cal G}_{ij}(E)$ and hence the resulting spectrum is
still particle-hole symmetric. In Fig.\ref{TP3}, we show the
comparison of the spectrum for $H_I$ by using ${\cal G}_{ij} (E)$
and ${\cal G}_{ij} (E=0)$. It is clear that the difference is
small and both spectra follow the Porter-Thomas distribution, in
agreement with the derivation in the Appendix that is based on
saddle-point approximation.

\section{Effects of Coulomb interaction and ferromagnetism} \label{ferro}

In this section, we discuss effects of the Coulomb interaction due
to the change of density of states.  To include the effects of
Coulomb interaction, we note that for neutral graphene, even though
excitonic effects are expected to be large\cite{Louie}, for low
energies and long distances, screening can be taken into
consideration by the renormalization of $v$ and the dielectric
constant $\epsilon$\cite{Sheehy}. Hence for low density of defects
in which separation between any two defects is large, one needs to
replace $v$ by $v_R$ in Eqs. (\ref{GAAr}) and (\ref{GABr}). This
would effectively replace the hopping amplitude $t$ by $t_R$.

\begin{figure}[htbp]
\hspace{-0.2in}
\includegraphics[width=5.5cm]{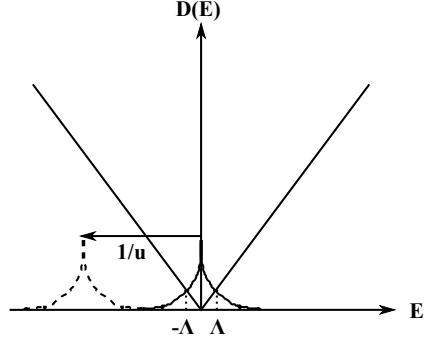}
\caption{Schematic plot of the impurity band and the original Dirac
band. Here the solid lines at center represent the impurity band
when $u = \infty$. For a small finite $u$, the impurity band
(represented by the dash line) is shifted into the Dirac band and
disappears} \label{states}
\end{figure}

As indicated in the introduction, strong disorders in a graphene
are a possible source for the observed ferromagnetism. To examine
whether the Porter-Thomas type distribution supports
ferromagnetism, we first note that the normalization adopted in
Eq.(\ref{Dlambda}) has to be corrected by taking into account the
conservation of states. As indicated in Fig.\ref{states}, since
resonant states replaces states in Dirac band, it requires a
cutoff $\Lambda$ in the impurity band so that numbers of states
for the impurity band and the Dirac band are
equal. Since number of states for the impurity band per site is
$n_I$, integration of the Dirac band yields $\Lambda = \sqrt{\pi/4}
v_R n_I$. By including the cutoff $\Lambda$, appropriate normalized
${\cal D}_{\Lambda}$ is given by
\begin{equation}
{\cal D}_{\Lambda} (\lambda ) = \frac{n_I}{\sqrt{8 \pi \langle
|\lambda|\rangle} {\rm erf} (\sqrt{\lambda_{\Lambda}/2\langle
|\lambda|\rangle})} e^{- \frac{|\lambda|}{2 \langle
|\lambda|\rangle} }. \label{Dlambda1}
\end{equation}
Here erf is the error function and $\lambda_{\Lambda} = \lambda
(\Lambda)$. Note that when $v$ is renormalized to $v_R$, both
$\langle |\lambda| \rangle$ and $\lambda_{\Lambda}$ are renormalized
by the same factor $v_R/v$.

To investigate the magnetism, we note that the electron wavefunction
$\psi_E$ is related to the eigenfunction $\phi$ of $H_I$ as
follows\cite{mou09}
\begin{equation}
\psi_E (\vec{r}_i) = \sum_{j = {\rm defect \hspace{1pt} positions}}
{\cal G}_{ij} A^j_E. \label{psiphi}
\end{equation}
Here $A^j_E$ is proportional to $\phi_j$. The normalization of
$A^j_E$ is determined by $\langle \sum_i \psi^2_E (\vec{r}_i)
\rangle =1$. The applicability of the Porter-Thomas distribution
implies that $\phi_j$ (thus $A^j_E$) follows Gaussian
statistics\cite{ThomasPorter}. Hence we have
\begin{equation}
\langle A^i_E A^j_E \rangle = \Gamma \delta_{ij}.
\end{equation}
By expressing ${\cal G}_{ij} = \frac{1}{M/2} \sum_{\vec{q}} {\cal G}
(\vec{q}) e^{i\vec{q} \cdot (\vec{r}_i-\vec{r}_j)} $ and using the
fact that ${\cal G}_{AA}(r,E=0) = {\cal G}_{BB}(r,E=0) =0$, we find
$\Gamma = 1/(N_I \gamma)$ with
\begin{equation}
\gamma = \frac{1}{ M/2} \sum_{\vec{q}} {\cal G}_{AB}(\vec{q}){\cal
G}_{AB}(-\vec{q}) .
\end{equation}
We shall include the Coulomb interaction by calculating the
exchange energy. For a neutral graphene, it is known that screened
Coulomb interaction is still long-ranged\cite{Sheehy, Geim07}
\begin{equation}
H_C = \frac{e^2}{8 \pi \epsilon} \sum\limits_{i,j,\sigma,\sigma'}
C^{\dagger}_{i \sigma} C_{i \sigma} \frac{1}{|\vec{r}_i -\vec{r}_j|}
C^{\dagger}_{j \sigma'} C_{j \sigma'}, \label{H_C}
\end{equation}
where $\epsilon$ is the renormalized dielectric constant and is
roughly $2.3 \epsilon_0$.  To obtain the exchange energy,
Eq.(\ref{psiphi}) is replaced by $C^{\dagger}_{i \sigma} =
\sum_{E,j} A^j_E {\cal G}_{ij} (E)C^{\dagger}_{E \sigma}$. By
setting any pair of $C^{\dagger}_{E \sigma} C_{E' \sigma'}$ by its
average value $\langle C^{\dagger}_{E \sigma} C_{E' \sigma'} \rangle
$, using the fact $\langle A^i_{E_1} A^j_{E_2} A^k_{E_2} A^l_{E_1}
\rangle  = \langle A^i_{E_1} A^l_{E_1} \rangle \langle A^j_{E_2}
A^k_{E_2} \rangle $ and approximating ${\cal G}_{ij} (E)$ by ${\cal
G}_{ij} (0)$, we find that the exchange energy is given by
\begin{equation}
E_{ex} = - \frac{e^2(n^2_{\uparrow}+n^2_{\downarrow})}{8 \pi
\epsilon \gamma^2} B \label{exchange}
\end{equation}
with
\begin{equation}
B=\sum\limits_{i,j} \frac{1}{|\vec{r}_i -\vec{r}_j|} \left(
\sum\limits^{N_I}_{k=1} {\cal G}_{ik} {\cal G}_{jk} \right)^2,
\end{equation}
 where $n_{\sigma} = N_{\sigma} /N_I$ are fractions of electrons
in the spin state $\sigma$. By approximating ${\cal G}_{ij} (E)$ by
${\cal G}_{ij} (0)$ and expressing ${\cal G}_{ij}$ in Fourier space,
we find
\begin{eqnarray}
&& \frac{B}{M} = \frac{16 \pi n^2_I}{\sqrt{3} a^2 M^2}\times
\nonumber \\
&& \sum_{q,q'} \frac{1}{|\vec{q}+\vec{q}'|} {\cal
G}_{AB}(\vec{q}){\cal G}_{AB}(-\vec{q}) {\cal G}_{AB}(\vec{q}'){\cal
G}_{AB}(-\vec{q}') \label{B}
\end{eqnarray}
where only the $i$ and $j$ in the same sublattice would
contribute. Since for $E \sim 0$, Eq.(\ref{GBA}) implies ${\cal
G}_{AB}$ diverges near Dirac points $\vec{q}_D$. The main
contribution in the integral of $B$ comes from regions of
$\vec{q}_D$. By setting $\vec{q}$ to any one of the Dirac points
in the factor $1/|\vec{q}+\vec{q}'|$, we find
\begin{equation}
\frac{B}{M} = \frac{2+\sqrt{3}}{4a} n^2_I \gamma^2. \label{BM}
\end{equation}

In the ferromagnetic state, we have $n_{\uparrow} \neq
n_{\downarrow}$ with $E_{\sigma}$ being the corresponding Fermi
energy for the spin state $\sigma$. The net spin moment is
proportional to $ m \equiv n_{\uparrow}-n_{\downarrow}$.
Substituting Eq.(\ref{BM}) back to $E_{ex}$, we find that the
exchange energy per site due to $m$ is given by
\begin{equation}
\frac{E_{ex}}{M} = - \frac{e^2 m^2 }{16 \pi \epsilon}
\frac{2+\sqrt{3}}{4a} n^2_I. \label{EexM}
\end{equation}
\begin{figure}[htbp]
\hspace{-0.2in}
\includegraphics[width=7.0cm]{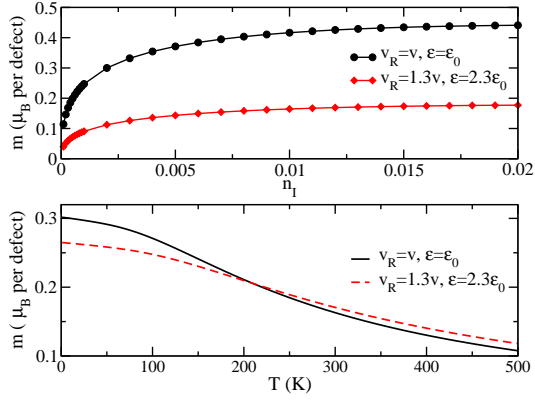}
\caption{(Color on-line) Magnetization per defect for $u = \infty$
with screened ($\epsilon=2.3 \epsilon_0$, $v_R=1.3v$) and unscreened
($\epsilon =\epsilon_0$, $v_R=v$) Coulomb interactions. (a)
Magnetization versus defect density at zero temperature (b) A
typical temperature dependence of magnetization for
$n_I=O(10^{-3})$. There is no sharp transition temperature. However,
by linear extrapolation, one finds that $T_c$ is around $600-700K$.}
\label{magnetization}
\end{figure}
For an undoped graphene, $E_{\downarrow}=-E_{\uparrow}$. In this
case, the net spin $m$ can be expressed as  $ m =  2
\int^{E_{\uparrow}}_0 dE D_{\Lambda} (E) $, while the change of the
total energy in the impurity band per site is $\Delta k = 2
\int^{E_{\uparrow}}_0 dE E D_{\Lambda}(E)$. The minimization of
$\Delta k + E_{ex}/M$ with respect to $E^{\uparrow}$ then leads to
\begin{equation}
\frac{(2+\sqrt{3}) e^2 n_I }{16 \pi \epsilon a} \int^{\lambda
({E_{\uparrow})}}_0 {\cal D}_{\Lambda} (\lambda) d \lambda =
E_{\uparrow}. \label{EexM}
\end{equation}
Solving Eq.(\ref{EexM}) yields $E_{\uparrow}$ which in turn
determines the magnetization per defect at zero temperature. The
same calculation can be easily generalized to any finite
temperature $T$. In this case, the magnetization is still
determined by the minimization of $\Delta k + E_{ex}/M$ with
respect to $E^{\uparrow}$ except that now $m$ and $\Delta k$ are
replaced by
\begin{eqnarray}
 m =  \int^{\infty}_{-\infty} dE D_{\Lambda} (E) \left[
n_{\uparrow} (E)
- n_{\downarrow} (E) \right], \\
\Delta k = \int^{\infty}_{-\infty} dE D_{\Lambda} (E) \left[
n_{\uparrow} (E) + n_{\downarrow} (E) -2 n_0(E) \right],
\end{eqnarray}
where $n_{\sigma} (E)= 1/(e^{\beta(E+E_{\sigma})}+1)$ are the
Fermi-Dirac distributions for $\sigma = \uparrow$ or $\downarrow$
and $n_0 (E)= 1/(e^{\beta E}+1)$ with $\beta =1/k_B T$.

In Fig.\ref{magnetization}(a), we show the magnetization at zero
temperature for $u=\infty$ with screened and unscreened Coulomb
interaction by solving Eq.(\ref{EexM}). It is seen that screening
reduces the magnetization. Furthermore, due to the divergent DOS at
$E=0$, ferromagnetism persists down to zero defect density and
magnetization increases as defect density increases.
Fig.\ref{magnetization}(b) shows a typical temperature dependence of
magnetization for $n_I=O(10^{-3})$. The temperature dependence shows
a quasi-linear behavior with a Boltzman tail. To compare with
experiments, we perform the linear extrapolation of magnetization
curve, which indicates that $T_c \sim 600-700 K$, in agreement with
experimental observations\cite{Barzola}.

\section{Summary and discussion} \label{conclude}

In summary, in this work we have shown that in the strong disorder
limit, a resonant impurity band is induced in a graphene. By
combining analytic arguments and numerical calculations, we show
that the density of resonant states is governed by the principle of
equal {\it a priori} probabilities for squared magnitudes of
eigenfunctions of a Euclidean Random Matrix. For large on-site
defect potential, the principle of equal {\it a priori}
probabilities shows that the density of resonant states is
characterized by the Thomas-Porter distribution and is divergent
near zero energy. Furthermore, we show that the observed
ferromagnetism is due to the combination of strong disorder and
long-range Coulomb interaction. The linear extrapolation of
magnetization curve indicates that $T_c \sim 600-700 K$, as observed
in experiments.

In addition to the magnetism, the impurity band enhances the
transport\cite{mou09}. This is consistent with experimental
observations\cite{Coleman} but is quite different from ordinary
impurity states even though in the calculated participation number
of $H_I$ in Fig.\ref{TP1}, some eigenfunctions $\phi$ are localized.
The crucial difference lies in the semi-localized nature of the
electronic states as revealed in Eq.(\ref{psiphi}). Here even though
$A^j_E$ (thus $\phi^j$) is localized, due to that ${\cal G} \sim
1/r$, $\psi_E$ will not be exponentially localized around defect
positions. The participation number for $\psi_E$ itself is of the
order of $(\ln M)^2$, indicating its semi-localized nature.

While so far in this work we only consider the strong disorder
limit, the results also provide some insight into the weak
disorder region. As illustrated in Fig.\ref{states}, for weak
disorders, $u$ is small, the impurity band is shifted into the
Dirac band. In this case, while the majority weight of the
impurity band disappears, its tail still sweeps through zero
energy and contributes small but finite DOS. As indicated above,
these density of states generally enhances the transport. This
explains why when graphene is made cleaner, the conductivity,
instead of increasing, decreases and appears to approach an
universal constant\cite{Geim07}. While the impurity band cannot
account for the exact value of the universal conductivity, our
results serve as a useful starting point for obtaining corrections
to the conductivity.

\begin{acknowledgments}
We thank Profs. T. K. Ng, Ting-Kuo Lee, and Hsiu-Hau Lin for
discussions. This work was supported by the National Science Council
of Taiwan.
\end{acknowledgments}

\appendix

\section{Saddle-point limit and Porter-Thomas distribution}

In this appendix, we shall show that the eigenvalue distribution for
$H_I$ follows the Porter-Thomas distribution in the limit of $N_I
\rightarrow \infty$ but with the defect density $n_I$ being fixed.
We start by noting that the spectrum of $H_I$ can be found by
calculating the resolvent
\begin{equation}
R(z) = \langle \frac{1}{M} Tr \frac{1}{z-H_I} \rangle,
\end{equation}
where $M$ is the total number of lattice points and $\langle \cdot
\rangle$ is the average over the random configurations of defects.
Clearly, we have $\cal{D} \it{(\lambda ) = - \frac{1}{\pi} {\rm
Im} R (\lambda + i 0^+)}$. As shown in the text, since the
spectrum has particle-hole symmetry, we shall set $\lambda$ to
$|\lambda|$ and consider only positive $\lambda$. The evaluation
of $R(z)$ can be reformulated by a replica field theory\cite{Zee}
via the following identity
\begin{eqnarray}
R(z) & = & \lim_{n \rightarrow 0} \frac{-1}{nM}
\frac{\partial}{\partial
z} \langle e^{-n Tr \log (z-H_I) } \rangle \nonumber \\
&=& \lim_{n \rightarrow 0} \frac{-1}{nM} \frac{\partial}{\partial z}
\left\langle \frac{1}{\det(z-H_I)^n} \right\rangle. \label{R}
\end{eqnarray}
The term $1/ \det (z-H_I)^n$ can be re-expressed by $n$ replica
complex fields $\phi_a$ ($a$=1,2,3,..,$n$) as follows
\begin{eqnarray}
\left\langle \frac{1}{\det(z-H_I)^n} \right\rangle  = \left\langle
\int \prod\limits^{N_I}_{i=1} \prod\limits^{n}_{a=1} {\cal D} \phi
e^{-\sum_{ij} {\phi^a_i}^* (H_{I})_{ij} \phi^a_j} \right\rangle . \nonumber \\
\label{replica}
\end{eqnarray}
Up to now $\phi_a$ is only defined on defect sites. To remove this
constraint, one introduces the field $\hat{\psi}_a$ defined on every
lattice site and impose $\delta [ \hat{\psi}_a (\vec{r}) - \sum_i
\phi^a_i \delta_{\vec{r},\vec{r}_i} ]$. The constraint of the delta
function can be removed by using the identity $\delta (F) = \int d
\psi_a e^{i\psi_a F}$. Here $\psi_a$ is the replica field. After
integrating out $\phi^a_i$ and $\hat{\psi}_a$, the resolvent can be
expressed as\cite{Zee}
\begin{equation}
R(z)= - \lim_{n \rightarrow 0} \frac{1}{nM} \frac{\partial \log
Z}{\partial z},
\end{equation}
where the partition function $Z$ is given by $Z=\int {\cal D} \psi
e^{-S} $ with $S$ being given by
\begin{equation}
S=- \sum^n_{a=1} \sum_{i,j} \psi^*_a (\vec{r}_i) F_{ij} \psi_a
(\vec{r}_j) -n_I \sum_i e^{-\frac{1}{z} \sum_a |\psi_a
(\vec{r}_i)|^2} .
\end{equation}
Here $F_{ij} = {\cal G}^{-1}_{ij} - \delta_{ij} {\cal G}^{-1}_{ii}$.
Note that because $G^0 = 1/(E-H)$, we have that for $E \sim 0$,
$F_{ij}= -(H_0)_{ij}$. In other words, $-F$ has the same form as the
tight-binding Hamiltonian for graphene except that it only acts on
defect sites.

After substituting $Z$ back to $R$, we find
\begin{widetext}
\begin{eqnarray}
 R(z) = - \lim_{n
\rightarrow 0} \frac{n_I}{nM} \frac{\int {\cal D} \psi \frac{1}{z^2}
\sum_{a,i} |\psi_a (\vec{r}_i)|^2 e^{-\frac{1}{z} \sum_b |\psi_b
(\vec{r}_i)|^2} e^{-S} }{\int {\cal D} \psi e^{-S}} \nonumber \\ = -
\lim_{n \rightarrow 0} \frac{n_I}{M }  \frac{\int {\cal D} \psi
\frac{1}{z^2} \sum_i |\psi_1 (\vec{r}_i)|^2 e^{-\frac{1}{z} \sum_b
|\psi_b (\vec{r}_i)|^2} e^{-S} }{\int {\cal D} \psi e^{-S}}.
\nonumber \\ \label{R1}
\end{eqnarray}
\end{widetext}
Here we have made use of the equivalence among different replica
component $a$ and the equivalence among different positions
$\vec{r}_i$ in the second equality.

It is clear that in Eq.(\ref{R1}), the replica symmetry is broken.
One needs to perform integrations for $\psi_1$ and $\psi_a$ with $a
\neq 1$ separately. For $\psi_1$, the integrand can be rewritten as
$\sum_i e^{-S^i_1}$ with $S^i_1$ given by
\begin{equation}
S^i_1= \frac{|\psi_1 (\vec{r}_i) |^2}{z} - \ln |\psi_1 (\vec{r}_i)
|^2 +S.
\end{equation}
Since we shall be interested in $z \sim 0$, i.e., energy near zero,
in the saddle-point approximation, integration over $\psi_1$ is
dominated by the maximum of $S^i_1$, which is determined by $
\frac{\partial}{\partial \psi^*_1(\vec{r}_i) } S^i_1 =0$ for all
$\vec{r}_i$. We find that maximum of $S^i_1$ satisfies
\begin{eqnarray}
& & \frac{|\psi^0_1 (\vec{r}_i) |^2}{z}( 1+ n_I e^{-\frac{1}{z}
\sum_a |\psi_a (\vec{r}_i) |^2} ) -1 \nonumber \\ & &- \psi^{0*}_1
(\vec{r}_i) \sum_j F_{ij} \psi^0_1 (\vec{r}_j) =0. \label{max}
\end{eqnarray}
It is clear that for low density, we can expand $\psi^0_1$ in term
of $n_I$. We find $|\psi^0_1 (\vec{r}_i) |^2 / z = 1- n_I
e^{-\frac{1}{z} \sum_a |\psi_a (\vec{r}_i) |^2} -\psi^{0*}_1
(\vec{r}_i) \sum_j F_{ij} \psi^0_1 (\vec{r}_j) + \cdots$. Because
$F_{ij}$ is finite, we obtain $\psi^0_1 (\vec{r}_i) \sim \sqrt{z}$
for all $\vec{r}_i$. As a result, the integration of $\psi_1$ in
Eq.(\ref{R1}) can be approximated as
\begin{eqnarray}
& & \frac{1}{M} \int {\cal D} \psi_1 \frac{1}{z^2} \sum_i |\psi_1
(\vec{r}_i)|^2 e^{-\frac{1}{z} \sum_b |\psi_b (\vec{r}_i)|^2} e^{-S}
\nonumber \\
 & & \sim \frac{1}{M} \sum_i \sqrt{\frac{2 \pi}{(S^{i}_1)^{''}}} \frac{z}{z^2}
e^{-zb} I_i \sim \frac{1}{M} \sum_i \frac{e^{-bz}}{\sqrt{z}} I_i.
\label{a1}
\end{eqnarray}
Here $b$ is a constant that results from $F_{ij}$ and
$(S^{i}_1)^{''}$ is the 2nd derivative of $S^i_1$ with respect to
$\psi^0_1$ and is proportional to $1/z$. $I_i$ is the integration
over $\psi_a( \vec{r}_i)$ with $a \neq 1$ and is given by
\begin{eqnarray}
I_i &=& \int {\cal D} \psi_a ({\vec{r}_i}) e^{-\frac{1}{z} \sum_{a
\neq 1}
|\psi_a (\vec{r}_i)|^2} e^{-S'} \nonumber \\
&=& \langle  e^{-\frac{1}{z} \sum_{a \neq 1} |\psi_a (\vec{r}_i)|^2}
\rangle' Z'_{n-1},
\end{eqnarray}
where $\langle \rangle'$ is the average with respect to $S'$ and
$Z'_{n-1} = \int {\cal D} \psi_a e^{-S'}$ with $S'$ being given by
\begin{equation}
S'=- \sum^n_{a=2} \sum_{i,j} \psi^*_a (\vec{r}_i) F_{ij} \psi_a
(\vec{r}_j) -n_I e^{-1} \sum_i e^{-\frac{1}{z} \sum_a |\psi_a
(\vec{r}_i)|^2} .
\end{equation}
It is clear that different $a$ and $i$ are equivalent. Therefore, we
obtain
\begin{equation}
\frac{1}{M} \sum_i I_i = \langle e^{-\frac{n-1}{z} |\psi_b
(\vec{r}_i)|^2} \rangle' Z'_{n-1}. \label{a}
\end{equation}
Combing Eqs. (\ref{a1}) and (\ref{a}) gives the limiting behavior
of the numerator for small $z$. To obtain the spectrum for small
$\lambda$, one needs to find the analytical continuation by
replacing $z$ by $\lambda+i0^+$. Clearly, the factor
$e^{-bz}/\sqrt{z}$ only contributes the real part. Together with
the fact that the denominator is $Z = e^{-n Tr \log (z-H_I)}$,
which goes to one when $n$ approaches zero, we find that
\begin{equation}
{\cal D} (\lambda) = n_I \frac{e^{-b \lambda}}{\sqrt{\lambda}}
\lim_{\lambda \rightarrow 0, n \rightarrow 0} {\rm Im} \langle
e^{\frac{|\psi_b (\vec{r}_i)|^2}{\lambda + i 0^+} } \rangle'
Z'_{-1}.
\end{equation}
Both $\langle e^{\frac{1}{z} |\psi_b (\vec{r}_i)|^2} \rangle'$ and
$Z'_{-1}$ can be calculated perturbatively with finite results in
the limit $\lambda \rightarrow 0$\cite{Zee}. After appropriate
normalization, one finds the spectrum of $\lambda$ follows the form
\begin{equation}
{\cal D} (\lambda) = n_I e^{-b |\lambda|} \sqrt{ \frac{b}{4 \pi
|\lambda|}}.
\end{equation}

\end{document}